\documentclass{aa}

\usepackage{graphicx}
\usepackage{txfonts}
\usepackage{color}
\usepackage{xspace}
\usepackage{amsmath}
\usepackage{amssymb}
\usepackage[hyperref, dvipsnames]{xcolor}
\usepackage{dashbox}
\usepackage{framed}
\usepackage{lipsum}
\usepackage{placeins}
\usepackage{stfloats}
\usepackage{hhline}
\usepackage{pifont}
\usepackage{tcolorbox}
\usepackage{placeins}

\bibpunct{(}{)}{;}{a}{}{,}

\usepackage{hyperref}

\hypersetup{
  colorlinks=true,
  breaklinks=true,
  linkcolor=blue,
  citecolor=blue,
  filecolor=blue,
  urlcolor=blue,
  unicode=false,
  pdftoolbar=true,
  pdfmenubar=true,
  pdffitwindow=false,
  pdfstartview={Fit},
  pdftitle={NEMESIS - Linking Young Stellar Object Morphology to Evolutionary Stages with Self-Organizing Maps},
  pdfauthor={David Hernandez},
  pdfsubject={YSO morphology and evolutionary stage},
  pdfcreator={David Hernandez},
  pdfkeywords={},
  pdfnewwindow=true,
  pdfdisplaydoctitle=true
}

\makeatletter
\renewcommand*\aa@pageof{, page \thepage{} of \pageref*{LastPage}}
\makeatother

\begin{document}

\title{Linking Young Stellar Object Morphology to Evolutionary Stages with Self-Organizing Maps}

\author{
David~Hernandez\inst{\ref{inst:uni_vie}} \and
Odysseas~Dionatos\inst{\ref{inst:nhm_vie}} \and
Marc~Audard\inst{\ref{inst:uni_ge}} \and
G\'{a}bor~Marton\inst{\ref{inst:konkoly},\ref{inst:mtaexcellence}} \and
Julia~Roquette\inst{\ref{inst:uni_ge}} \and
Ilknur~Gezer\inst{\ref{inst:konkoly},\ref{inst:mtaexcellence}} \and
Máté~Madarász\inst{\ref{inst:konkoly},\ref{inst:mtaexcellence},\ref{inst:uni_szeg}} \and
Kai~L.~Polsterer\inst{\ref{inst:hits}} 
}

\institute{
Universität Wien, Institut für Astrophysik, T\"urkenschanzstrasse 17, 1180 Wien, Austria\label{inst:uni_vie}
\\ \email{david.hernandez@univie.ac.at} \and
Natural History Museum Vienna, Burgring 7, 1010 Vienna, Austria\label{inst:nhm_vie} \and
Universit\'e de Gen\`eve, Department of Astronomy, Chemin Pegasi 51, 1290 Versoix, Switzerland\label{inst:uni_ge} \and
Konkoly Observatory, Research Centre for Astronomy and Earth Sciences, Hungarian Research Network (HUN-REN), Konkoly Thege Mikl\'{o}s \'{U}t 15-17, 1121 Budapest, Hungary\label{inst:konkoly} \and
CSFK, MTA Centre of Excellence, Budapest, Konkoly Thege Miklós út 15-17., H-1121, Hungary\label{inst:mtaexcellence} \and
HITS gGmbH, Astroinformatics, Schloss-Wolfsbrunnenweg 35, 69118 Heidelberg, Germany\label{inst:hits} \and
Department of Experimental Physics, Institute of Physics, University of Szeged, D{\'o}m t{\'e}r 9, 6720 Szeged, Hungary\label{inst:uni_szeg}
}

\date{Received 12 December 2024 / Accepted 22 September 2025}

\abstract{
  Many studies over the past few decades have investigated young stellar object evolution based on their spectral energy distribution (SED). The SED is heavily influenced not only by evolutionary stage, but also the morphology of the forming star. This study is part of the NEMESIS project which is aiming to revisit star formation with the aid of machine learning techniques and provides the framework for this work.
}
{
  In a first effort towards a novel spectro-morphological classification we analyzed young stellar object morphologies and linked them to the currently used observational classes. Thereby we aim to lay the foundation for a spectro-morphological classification, and apply the insights learned in this study in a future, revisited classification scheme.
}
{
  We obtained archival high-resolution survey images from VISTA for approximately \(10\,000\) literature young stellar object candidates towards the Orion star formation complex (OSFC). Utilizing a Self-Organizing map (SOM) algorithm, an unsupervised machine learning method, we created a grid of morphological prototypes from near- and mid-infrared images. Furthermore, we determined which prototypes are most representative of the different observational classes, derived from the infrared spectral index, via Bayesian inference.
}
{
  We present our grids of morphological prototypes of young stellar objects in the near-infrared, which were created purely from observational data. They are thus nondependent on theoretical models. In addition, we show maps that indicate the probability for a prototype belonging to any of the observational classes.
}
{
  We find that SOMs created from near-infrared images are a useful tool, with limitations, to identify characteristic morphologies of young stellar objects in different evolutionary stages. This first step lays the foundation for a spectro-morphological classification of young stellar objects to be developed in the future.
}


\maketitle

\section{Introduction}
\label{sec:introduction}%
When new stars are born, they go through several evolutionary stages: from the very beginning when the source material in a molecular cloud collapses under its gravity up until the point when hydrogen fusion in the core is self-sustained by the star \citep[see, e.g.,][]{lada1984,lada1987,andre1993,palla1996,whitney2003a,robitaille2006}. Objects that are in these early stages of star formation are known as young stellar objects (YSOs). To better understand the star formation process, it is crucial to identify the timescales and order of the different evolutionary stages \citep[see, e.g.][]{lada2003,williams2011, morbidelli2016, drazkowska2023}.

Extensive work to develop and refine methods that attempt to estimate the current evolutionary stage of a YSO have been developed in the past few decades \citep[see, e.g.,][]{gutermuth2009,evans2009,rebull2010,megeath2012,koenig2014,dunham2015}. A common approach to estimate the current evolutionary stage of a YSO analyses the source's spectral energy distribution (SED) to derive an observational class. The current standard classification for YSOs is based on the infrared spectral index \(\alpha_\mathrm{IR}\), which was first defined in the range from \(2.2\) to \(25\;\mu\mathrm{m}\) by \citet{lada1987}. The original Lada classification had three distinct observational classes -- I, II, and III -- which correspond to different ranges of the \(\alpha_\mathrm{IR}\)-index, i.e., they are proxies to the shape of the infrared SED \citep{lada1987}.

Since then, the Lada classification has undergone several revisions and updates, most notably with the discovery of the deeply embedded, later dubbed Class 0 YSOs \citep{andre1993} as a precursor to Class I YSOs. Unfortunately, these youngest protostars are difficult to observe in wavelengths shorter than the far infrared due to the protostellar envelope absorbing the emission from the central source. Thus, it poses challenging to separate these Class 0 YSOs from the more evolved Class I objects. In addition to the criterion derived by \citet{andre1993} based on the ratio of sub-millimeter and bolometric luminosity -- i.e., \(L_\mathrm{submm}/L_\mathrm{bol}\) -- the bolometric temperature, \(T_\mathrm{bol}\), was introduced as an alternative to the \(\alpha_\mathrm{IR}\)-index \citep{myers1993}. \citet{chen1995} further refined this system by defining four classes -- 0 to III -- of YSOs based on the bolometric temperature of their observed SEDs. However, it is necessary to have a well sampled SED to calculate \(T_\mathrm{bol}\) with confidence.

Furthermore, a flat-spectrum class \citep{greene1994} was introduced for those sources that exhibit a flat slope along the infrared SED. The true nature of these flat-spectrum sources is yet unknown. However, a few theories are discussed in the literature. One explanation is given e.g., by \citet{greene2002}, suggesting that flat-spectrum sources are an intermediate evolutionary stage between classes I and II. On the other hand, radiative transfer modeling of YSOs shows that the observed SED shapes of flat-spectrum sources can also be the result of the system's spatial orientation and morphology, such as disk inclination and disk flaring \citep[see, e.g.,][]{whitney2003b,whitney2003a,robitaille2006,robitaille2007}.

Unfortunately, relying on estimates of the true evolutionary stage by use of the observational classes based on the shape of the SED is ambiguous, since the SED is a degenerate representation of a three dimensional complex object. Photometry reduces any spatial information of the source in question into a singular value, namely the flux density of the source at the wavelength range defined by the chosen filter. One consequence is, as has been shown in several recent studies, that the shape of the SED strongly depends on the angle at which a morphological complex YSO is observed, which by extension can lead to misclassification of the same \citep[see, e.g.,][]{crapsi2008,furlan2016,sheehan2022}.

Moreover, the youngest YSOs, for instance, feature an SED that peaks in the far-infrared regime but are invisible in the mid- and near-infrared. This reddening of the SED is mainly caused by two different cases. First through extinction caused by the interstellar medium with the main contributor to this extinction being the molecular cloud the YSO is associated to \citep[e.g.,][]{mcclure2010}. In the second case the YSO is deeply embedded in its proto stellar envelope and self extinction is the main cause of the SED reddening. In this case the young sources actively drive accretion outflows impacting the surrounding envelope \citep{andre1993}. Unfortunately it is also possible to have a combination the two cases, where a deeply embedded YSO lies on the far side of its hosting molecular cloud. Without any further knowledge, such as the exact distance to the source, or the exact spatial distribution of the molecular cloud it is not possible to compensate for any misclassification based on the SED.

However, it is possible to directly observe the geometry of an embedded YSO, as light from the central source is scattered on the walls of the outflow cavity inside the surrounding envelope. The scattered emission from the cavity walls is observable in the near-infrared (NIR) wavelengths as a uni- or bipolar cone with the narrow ends pointing towards the obscured central source \citep{kenyon1993,padgett1999,habel2021}. This suggests that, the spatial structures observed in images of YSOs are a direct consequence of the star formation process at our current understanding. For instance, stellar rotations observed in T Tauri stars are much slower than what was expected from a collapsing cloud of gas and dust. A possible answer, crediting for the missing angular momentum in Class II YSOs, is the formation of jets and molecular outflows, which are feedback mechanisms that can remove excess angular momentum from the system \citep[][]{shu1987, ray2021}. This begs the question if we can find a correlation between the observed morphology and the evolutionary stage of a YSO.

In a recent paper, \citet{habel2021} investigated how the protostellar envelope evolves under the influence of accretion feedback in young protostars in Orion. Their study is based on a carefully selected sample of 304 YSOs, consisting of Class 0, Class I and flat-spectrum sources, which were then compared to a grid of model images that are based on theoretical models of YSOs. \citet{habel2021} combined radiative transfer codes from \citet{whitney1992} and \citet{whitney1993} with model assumptions for the protostellar envelope from \citet{terebey1984} and compared the model images to real NIR images obtained with Hubble. They found that there is a correlation between the observational class obtained from the SED slope and the spatial features seen in high resolution images.

In this work we expand on the work of \citet{habel2021} and applied unsupervised machine learning to a larger sample of \(\approx 10\,000\) bona fide YSOs of all observational classes in Orion to explore the possible correlation between YSO morphology and evolutionary stage. By using the \(\alpha_\mathrm{IR}\)-index calculated from archival photometric data as a proxy for the true evolutionary stage, we were able to conduct our analysis in a purely data driven way, independent of theoretical model calculations. With the insights gained from this research we intend to lay the foundation for a future revised and improved spectro-morphological classification scheme for YSOs that is more resilient to some, hopefully all weaknesses of the current YSO classification. This study is part of the Novel Evolutionary Model for the Early Stages of stars with Intelligent Systems (NEMESIS) project\footnote{\url{https://nemesis.konkoly.hu/}}, where we use machine learning techniques to revisit star formation in the age of big data.

In the following sections, we will describe how we obtained and processed the data (see Sect.~\ref{sec:data_preparation}) and the methods used to analyze and link YSO morphology to the current standard classification (Sect.~\ref{sec:methodology}). In Sect.~\ref{sec:experiments} we describe the experiments leading to a grid of morphological prototypes, which are presented in Sect.~\ref{sec:results}. The results and limitations of our method are then discussed in Sect.~\ref{sec:discussion}. The final Sect.~\ref{sec:conclusion} contains concluding remarks and a brief future outlook on how we plan to progress.

\section{Data preparation}
\label{sec:data_preparation}%
In the sections below we present the data used in this paper and describe how it was pre-processed for the \textbf{P}arallelized rotation and flipping \textbf{IN}variant \textbf{K}ohonen maps \citep[PINK;][]{polsterer2016}, a Self-Organizing map algorithm (SOM, see Sect.~\ref{sec:som}). Moreover, a homogeneous YSO classification for our training sample is needed to compare morphology to the current standard classification. 

\subsection{Data selection}
\label{sec:data_selection}%
To build the training sample for PINK, we selected all sources from the NEMESIS Catalogue of Young Stellar Objects for the Orion star formation complex (OSFC) compiled by \citet{roquette2025}. This catalog contains roughly \(27\,000\) sources that have been assigned the young stellar object label in the literature at some point since the 1980s. In this source catalog \citet[][]{roquette2025} have gathered all available data that is relevant for YSOs, such as photometric measurements to construct SEDs, astrophysical parameters such as \(T_\mathrm{eff}\) and \(L_\mathrm{bol}\), and equivalent widths of emission and absorption lines to name a few. From this data compilation, we use the photometric measurements to compute the observational class of the YSO.

To investigate different morphologies and associate them with observational classes, we require images of YSOs that show resolved structures associated with their host sources. Due to the high extinction environment, these young stars are native to, we resort to observations in the near- and mid-infrared wavelengths. The selected data contains observations in three photometric filters, utilizing archival data from the \emph{Visual and Infrared Survey Telescope for Astronomy} (VISTA) and the \emph{Spitzer} space telescope. The \emph{VISTA}-VIRCAM J, H, and K\textsubscript{s} passbands provide high resolution near-infrared (NIR) images. The areas highlighted by the green box are observed in the four IRAC filters, and the region enclosed by the red boundaries corresponds to the MIPS 1 passband.

\begin{figure*}
  \centering
  \includegraphics[width=17cm]{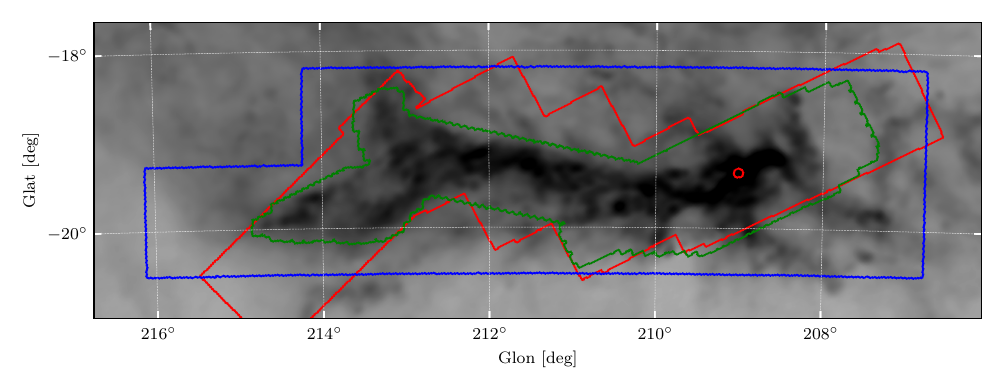}
  \caption{Survey footprints of the image data we used in this study. The blue outlines show the area covered by the J, H, and K\textsubscript{s} filters, the green box shows the area covered by the four IRAC filters, and the red lines highlight the area observed in the MIPS 1 filter. The background image shows the dust emission from the \emph{Planck} satellite at \(545\;\mathrm{GHz}\) \citep{planckcollaboration2020}.}
  \label{fig:survey_footprints}
\end{figure*}

Figure~\ref{fig:survey_footprints} shows the observation footprints for the image data in Orion that we used in this study. The blue outlines show the coverage of the OSFC in the NIR J, H, and K\textsubscript{s} filters. The VISION images cover an area of roughly \(18.2\,\mathrm{deg}^2\) \citep{meingast2016}. We obtained NIR images from the Vienna Survey in Orion \citep[VISION;][ESO program ID 090.C-0797(A)]{meingast2016} observation campaign\footnote{We received fully reduced mosaics from S. Meingast through private communication.}. The \emph{Spitzer} IRAC and MIPS 1 filters span an area of \(10.83\,\mathrm{deg}^2\) and \(16.93\,\mathrm{deg}^2\), respectively. In this study, we used \emph{Spitzer} data only to calculate the \(\alpha_\mathrm{IR}\)-index.

\subsection{Data extraction and preprocessing}%
\label{sec:data_extraction and preprocessing}

To find distinct morphological families of YSOs, we created post-stamp cutouts from the mosaic images centered on the sky coordinates of the sources in the NEMESIS YSO catalog \citep{roquette2025}. Since the main features we aimed to identify are the morphological structures surrounding the central objects, we had to choose the size of the image stamps appropriately. This entailed weighing a larger stamp size -- to include jets and outflows that have already traveled further from the source or are only visible at larger distances to the YSO -- against contamination from unrelated nearby sources and structures such as cloud emission from the surrounding interstellar medium (ISM). This is incredibly challenging in highly crowded areas like the Orion Nebula Cluster (ONC). We eventually converged toward a stamp size of \(50 \times 50\;\mathrm{arcsec}\), which at a mean distance of \(414\,\mathrm{pc}\) to the OSFC \citep{menten2007} corresponds to a square of approximately \(20\,000 \times 20\,000\;\mathrm{au}\). Image stamps of this size contain the protostellar disk \citep[size on the order of a few dozen to several hundred au;][]{tobin2020} and envelope \citep[can reach a size of several thousand au;][]{heimsoth2022}. They further partially cover jets and shocked material that emerge from the protostellar envelope, while keeping contaminating sources manageable. For our training set, we extracted \(\approx10\,000\) image cutouts per filter. The total number of post-stamp image cutouts gained from each passband is given in Table~\ref{tab:nr_of_images_per_filter}.

Downsampling the NIR images to a lower resolution has the benefit of shorter computation times when plugging them into our machine learning algorithm. As long as the new pixel scale stays below the seeing limit of the VISION survey, no spatial information is lost. Since the image resolution of VISION is limited by seeing -- which on average is \(0.78\arcsec\), \(0.75\arcsec\), and \(0.8\arcsec\) in the J, H, and K\textsubscript{s}-bands, respectively \citep{meingast2016} -- we opted to resample the images to a resolution of \(2\times10^{-4}\,\mathrm{deg}\,\mathrm{px}^{-1}\) which equals \(0.72\arcsec\mathrm{px}^{-1}\).

To clean the images in the training set from image artifacts caused by saturated sources, or sources that lie on the border of the image mosaic, we removed all images containing \verb+NaN+ values. In total a number of 492, 252, and 257 sources were removed from the sample in the J-, H-, and K\textsubscript{S}-band images, respectively. Table~\ref{tab:removed_sources} gives a detailed overview on the absolute and relative numbers of removed sources with respect to the total number in the training set. The table also lists in parenthesis the relative number of removed sources with respect to the number of sources in each class as determined by the \(\alpha_\mathrm{IR}\)-index (see Tab.~\ref{tab:yso_classes}).

\begin{table*}
  \centering
  \caption{Number of images removed from the training sample.}
  \label{tab:removed_sources}
  \begin{tabular}{l | r r | r r r | r r r | r r r}
    \hline\hline
    Class & \multicolumn{2}{c}{Total Nr. in training set} & \multicolumn{3}{c}{J-band removed} & \multicolumn{3}{c}{H-band removed} & \multicolumn{3}{c}{K\textsubscript{S}-band removed} \\
                   & Abs. & Rel. & Abs. & \multicolumn{2}{c |}{Rel.} & Abs. & \multicolumn{2}{c |}{Rel.} & Abs. & \multicolumn{2}{c}{Rel.} \\
    \hline
    0/I            &   723 &   6.98\% &  20 & 0.19\% & (2.77\%) &  13 & 0.13\% & (1.80\%) &  11 & 0.11\% & (1.52\%) \\
    flat spectrum  &   628 &   6.07\% &  39 & 0.38\% & (6.21\%) &  14 & 0.14\% & (2.23\%) &  16 & 0.16\% & (2.55\%) \\
    II             &  2967 &  28.65\% &  93 & 0.90\% & (3.13\%) &  56 & 0.54\% & (1.89\%) &  56 & 0.54\% & (1.89\%) \\
    III thin disk  &  2495 &  24.10\% & 119 & 1.15\% & (4.77\%) &  91 & 0.88\% & (3.65\%) &  96 & 0.93\% & (3.85\%) \\
    III no disk    &  1542 &  14.89\% &  81 & 0.78\% & (5.25\%) &  57 & 0.55\% & (3.70\%) &  46 & 0.44\% & (2.98\%) \\
    not classified &  2000 &  19.31\% & 140 & 1.35\% & (7.00\%) &  21 & 0.20\% & (1.05\%) &  32 & 0.31\% & (1.60\%) \\ \hline
    \(\Sigma\) & 10355 & 100.00\% & 492 & 4.75\% &  & 252 & 2.43\% &  & 257 & 2.48\% & \\
    \hline
  \end{tabular}
  \tablefoot{Percentages in parenthesis are with respect to the number of sources in that class.}
\end{table*}

\emph{Spitzer} images, albeit diffraction limited, have less resolution due to the smaller aperture of the telescope, and the larger pixel scale of the detectors. The IRAC 1, 2, 3, and 4 passband images have an average point spread function FWHM of \(1.66\arcsec\), \(1.72\arcsec\), \(1.88\arcsec\), and \(1.98\arcsec\), respectively\footnote{IRAC Instrument Handbook (\raggedright\url{https://irsa.ipac.caltech.edu/data/SPITZER/docs/irac/iracinstrumenthandbook/5/}) accessed in November 2024}. Furthermore, the MIPS 1 images have the least resolution with an FWHM of \(6\;\arcsec\)\footnote{MIPS Instrument Handbook (\raggedright\url{https://irsa.ipac.caltech.edu/data/SPITZER/docs/mips/mipsinstrumenthandbook/3/}) accessed in November 2024}.

For PINK to learn morphological prototypes, the YSO features -- i.e., structures directly related to the forming star, such as jets, outflows, and outflow cavities -- must be directly visible in the input images. Without further processing, these features are not necessarily immediately apparent as they may be hidden in the high dynamic range of the images. Thus, finding an optimal flux scaling that brings out these features and suppresses unrelated structures as best as possible is paramount.

To that end, we applied the stretching and normalization method developed by \citet{lupton2004} to the image cutouts. In general, the Lupton method allows stretching three filters together so that the color relations between the three channels are preserved, but single channel flux scaling is also possible. We opted to process each channel individually, especially since preliminary experiments with PINK had failed as the three channel mode had introduced processing artifacts that were confused for true features of YSOs by the SOM. We have experimented with other image pre-processing methods but with limited success. We found that automatically processing a large number of image cutouts containing different morphological structures is extremely challenging. For a morphological analysis alone our chosen pre-processing is sufficiently good, but for a future spectro-morphological classifier the flux scaling method should be revisited.

\section{Methodology}
\label{sec:methodology}%

In this section, we will describe the standard classification scheme for YSOs used as the classification baseline for the morphological analysis. More importantly, we will show how we created morphological prototypes of YSOs using the PINK Self-Organizing Map algorithm. Utilizing Bayesian inference, we explore how different observational classes link to the morphological prototypes on the SOMs. This enables us to identify typical source morphologies associated with certain classes of YSOs.

\subsection{YSO standard classification}%
\label{sec:standard_classification}

We first need homogeneously determined evolutionary classes for our YSO sample to link different YSO morphologies to an evolutionary stage. To that end, the classification scheme used by \citet{grossschedl2019}, which is based on the observational classes first defined by \citet{lada1987}, is adopted for our study. This method uses the shape of the infrared SED to estimate the true evolutionary stage and assigns an observational class accordingly. We can approximate the shape of the SED using the \(\alpha_\mathrm{IR}\)-index. To build a baseline of YSO classes for our training sample, we use non-linear least squares fitting to fit a power law in the range from 2 to \(24\;\mu\mathrm{m}\) to the infrared SED. We use all available photometric data available in that wavelength range for the fit.

With \(\alpha_\mathrm{IR}\) determined, we assign the observational class of the YSO using the \(\alpha_\mathrm{IR}\)-index ranges as reported in \citet{grossschedl2019}. By this definition, we have five distinct YSO classes: Class 0/I protostars, flat-spectrum sources, Class II T-Tauri stars, Class III sources with thin disk (anemic or debris disk), and diskless Class III pre-main sequence (PMS) or main sequence (MS) stars. Table~\ref{tab:yso_classes} lists the spectral index ranges reported in \citet{grossschedl2019}. Sources with insufficient data to compute the \(\alpha_\mathrm{IR}\)-index were not classified.

\begin{table}
  \caption{YSO classes and abundances in the training sample. The boundaries for the individual classes are adopted from \citet{grossschedl2019}.}%
  \label{tab:yso_classes}
  \centering
  \begin{tabular}{l c}
    \hline\hline 
    Class         & \(\alpha_\mathrm{IR}\) range         \\
    \hline
    0/I           & \(\hphantom{{-}0.3 \geq\;}\alpha_\mathrm{IR} > \hphantom{{-}}0.3\)   \\
    flat spectrum & \(\hphantom{{-}}0.3 \geq \alpha_\mathrm{IR} > -0.3\)                 \\
    II            & \(-0.3 \geq \alpha_\mathrm{IR} > -1.6\)                              \\
    III thin disk & \(-1.6 \geq \alpha_\mathrm{IR} > -2.5\)                              \\
    III no disk   & \(-2.5 \geq \alpha_\mathrm{IR}\hphantom{\;>{-}2.5}\)                 \\
    \hline
  \end{tabular}
\end{table}

\subsection{Self Organizing Maps}%
\label{sec:som}

SOMs are artificial neural networks \citep{kohonen1982} that are used to reduce the dimensionality of high-dimensional complex data \citep{kohonen2001}. In other words, a SOM provides a latent, low-dimensional, and discrete feature space that embeds the data, so that each data point from the input set can be characterized by its location and neighborhood in the latent space. We apply this method to a sample of NIR and MIR images (see Sect.~\ref{sec:data_selection}) of YSOs to build a grid of artificial image prototypes representing observed YSO morphologies in our training sample. These images can be represented as a vector where the vector dimension is equal to the number of pixels in the image. Thus, the general features of the high-dimensional data can be expressed through the coordinates of the low-dimensional latent space. In return, each location in the discrete latent space is equivalent to a high-dimensional image prototype, through which the embedding takes place based on the prototype's similarity to the input images. In our case where the input images show many morphological features of YSOs, the SOM creates a limited number of prototypes that show us how many different YSO morphologies were observed in the training set. This enabled us to quantitatively analyze YSO morphology without visually inspecting thousands of image cutouts one by one.

Moreover, a SOM is an unsupervised machine learning technique, that allows us to create the prototypes in a data-driven way without relying on a theoretical model calculation using, e.g., radiative transfer codes such as the ones used by \citet{robitaille2007}. The following paragraphs will give an overview of how a SOM works and introduce the \textbf{P}arallelized rotation and flipping \textbf{IN}variant \textbf{K}ohonen maps \citep[PINK;][]{polsterer2016} algorithm used in this study.

PINK is a SOM that creates rotation and flipping invariant prototypes for images of astronomical objects. In this paper, we use PINK to create a grid of exemplary images of YSOs which we then associate with their evolutionary stage via the \(\alpha_\mathrm{IR}\)-index. To create such a grid of images showing the morphological prototypes of YSOs, we initialize a cartesian grid of \(n\times m\) neurons in a random state, i.e., a noise image, and train the SOM in an iterative process that can be roughly divided into five steps.

\begin{enumerate}
  \item Draw a random element from the training sample.
  \item Create copies at different rotation and flipping states (Specific to PINK)
  \item Find the best matching unit (BMU).
  \item Compute the weights for all neurons in the map.
  \item Update the neurons in the map.
  \item Optional: Adapt the learning rate and neighborhood function.
\end{enumerate}

These steps are repeated until the SOM has settled. A SOM is considered settled either after the model has undergone a predefined number of epochs, or when the change of the SOM over several epochs stays below a certain threshold and the network is not learning new features \citep{kohonen2001}. To train our SOMs, we opted for a very simple training scheme.

\subsection{SOM training strategy and hyperparameters}%
\label{sec:hyperparameters}
For the best results, we have fine tuned the hyperparameters and divided the SOM training into two phases. We first populated the maps with a coarse version of possible morphologies by letting the algorithm run for \(1\,000\) epochs, choosing a broad neighborhood function and a large learning rate. The main learning phase for the SOM ran for \(15\,000\) epochs, but this time with a narrow neighborhood function and a small learning rate. We give additional information on how we chose the hyperparameters in Appendix~\ref{sec:appendix_training}. Table~\ref{tab:hyper_parameters} gives a brief overview of said hyperparameters.

\begin{table}
  \caption{Overview of PINK training hyperparameters.}
  \label{tab:hyper_parameters}
  \centering
  \begin{tabular}{l c c}
    \hline\hline
    parameter         & \multicolumn{2}{c}{value}                           \\\hline
    similarity metric & \multicolumn{2}{c}{Euclidean}                       \\
    neighborhood      & \multicolumn{2}{c}{Gaussian kernel}                 \\
    map size          & \multicolumn{2}{c}{20 by 20}                        \\
    map topology      & \multicolumn{2}{c}{cartesian non cyclic boundaries} \\\hline
                      & population phase  & learning phase                  \\\hline
    epochs            & \(1\,000\)        & \(15\,000\)                     \\
    kernel width      & 2.5               & 1.0                             \\
    learning rate     & 0.1               & 0.05                            \\
    no. of rotations  & 24                & 92
    \\\hline
  \end{tabular}
  \tablefoot{The top panel of the table shows the hyperparameters that are the same for both training phases. The bottom panel lists the hyperparameters that are unique to each of the two training phases.}
\end{table}

PINK can handle image rotation of up to \(1\degr\), i.e., 360 possible rotations (angles of freedom) per image. However, to save computation time, we limit the number of possible rotations to 24 and 92 for the first and second phases of training, respectively. Choosing fewer rotations for the first phase decreases the computation time but creates coarse prototypes. For smoother, more detailed neurons we increase the number of rotations to 92 for the main training phase, see Sect.~\ref{sec:training_the_maps}. Both training phases permit image flipping during training.

Furthermore, we chose cartesian noncyclic boundary topology for our map with a size of \(20\times 20\) neurons, see Appendix~\ref{sec:appendix_training} for further details on the chosen grid dimensions. In contrast to a cyclic boundary -- i.e., a doughnut shaped boundary, where the opposite edges and all corners of a square map connect -- the corners and edges of the SOM provide a space where distinct morphologies can retreat to and cluster together. This helps the algorithm to better separate different morphologies.

The map size of the SOM controls how sensitive the map is to differences in the morphology at a given abundance of expected morphological classes. For example, if \(10\%\) of the sources in the training sample exhibit morphologies other than point sources, and the SOM size is limited to 10 neurons, i.e., possible morphological prototypes, then we expect to have 9 neurons showing prototypes for point sources and only one prototype for all species of extended sources. Similar findings are reported by \citet[][see chapter 9 therein]{vantyghem2024}, along with possible remedies to counter the effect of underrepresented morphologies.

In our case, the extended YSOs, those in the main formation stages that we are investigating in this paper, make up roughly \(10\%\) of our training sample. To enable the SOM to also learn underrepresented morphologies we chose to optimize the number of neurons in the SOM, as is also suggested by \citet{vantyghem2024}. A SOM-size of \(20\times 20\) neurons has given us the best results without overfitting the model -- which would be the case if we had as many neurons as there are sources in our training sample -- resulting in one prototype for approximately 25 sources.

Finally, we created heatmaps to compare and find a relation between the YSO morphologies and the current standard classification, i.e., the \(\alpha_\mathrm{IR}\)-index. We used a Bayesian approach to map the individual YSO images from our training sample to the SOM prototypes. This provides us with a statistically sound framework, taking into account the estimated noise in each image when mapping the observations to the SOM. As a result, we are left with a probability distribution function on the SOM grid space for each filter.

\section{Experiments}
\label{sec:experiments}%
This section is dedicated to our experiments with PINK. We will give some details on our final training parameters and the filters we used to create the morphological prototypes. In addition, we show how we verify that the training of the maps has been successful.

\subsection{Training the maps}
\label{sec:training_the_maps}

Since we trained the maps for each filter individually, it is not guaranteed that similar morphologies will be found in the same locations on the maps for each filter if they are initialized randomly. That is, the prototypes found in the top left corner of the map for the J-band, may not be the same as those found in the top left of the map for any other passband. To keep the maps comparable across filters, only the J-band SOM was initialized in a random state. The SOMs for the longer wavelengths were then initialized with the fully trained SOM from the previous wavelength, i.e., the H-band SOM was initialized with the J-band SOM, and the K\textsubscript{s}-band SOM was initialized with the H-band SOM. We thereby ensure that we can, in general, find similar prototypes in similar regions across the SOMs for each filter.

Training of the maps is done in two steps; An initial map setting stage, and a main training stage. In the setting stage, we let the algorithm run for a \(1\,000\) epochs intended to populate the map with a first approximation of the different morphologies. Since at this stage, it is not too important to learn detailed YSO morphologies, we allow a limited number of 24 rotations plus flipping of the images to speed up the computation. Furthermore, we set the neighborhood function in this stage to a Gaussian kernel with a standard deviation \(\sigma\) of 2.5 neurons. The learning rate is set to 0.1 which gives the input images a higher weight when the map is updated also resulting in a faster population of the map.

The second stage allows 92 rotations for each input image, plus flipping, using a narrower kernel, and a lower learning rate. The number of rotations here yields a minimum rotation angle of \(3.91\degr\). Thus, at our chosen image size of \(69\times69\) pixels, rotating the cutout by the minimum angle about the image center shifts the middle pixel at the outer boundary of the image by \(\approx 2.5\) pixels. Since this is the main learning stage for the SOM, we have narrowed the width \(\sigma\) of the neighborhood function to 1 neuron in SOM coordinate space, dropped the learning rate to 0.05, and let the SOM train for \(15\,000\) epochs.

\subsection{Verifying training}

To verify that the maps have been trained well, we used PINKs mapping functionality to create heatmaps for each image in the training sample. A heatmap contains the Euclidean distance between the input image and each neuron in the SOM. Thus, we can use the heatmaps to extract the coordinates of the BMU, i.e., the neuron that is closest to the input image. We used the BMU coordinates, to count how many images from the training sample map to each neuron in the SOM. Furthermore, we used a histogram of BMU distances to have a second measure to assess the SOM training. In general, we require that two conditions are met; Firstly, we pursue that the mapped sources on the SOM are roughly uniform distributed, and secondly, we aim for small Euclidean distances to the BMU for the majority of sources in the training sample.

\begin{figure}
  \centering
  \resizebox{\hsize}{!}{\includegraphics{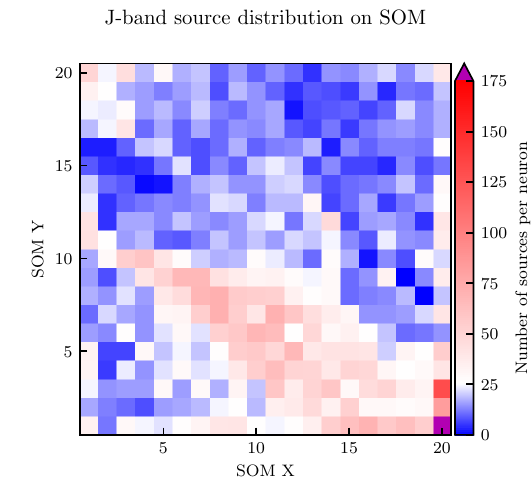}}
  \caption{Number of J-band image cutouts from the training sample mapping to the J-band SOM. Blue shades indicate regions on the SOM where fewer than the expected number of sources are mapped to, whereas red shades highlight regions where we find more sources than we expect after mapping them back to the SOM. White indicates the expected number assuming a uniform distribution.}
  \label{fig:j_band_source_distribution}
\end{figure}

For the first criterion in the J-band, we would ideally expect that each neuron is the best matching prototype for \(N_\mathrm{imgs~J-band}/N_\mathrm{prototypes}\) sources, which results in \(\approx25\) sources mapping to each neuron in this case.
Figure~\ref{fig:j_band_source_distribution} shows how the mapped J-band images distribute over the SOM prototypes. The color map in the plot has been chosen so that we can easily see any deviations above and below the estimated number of mappings per neuron. Even though we do not see a perfect homogeneous distribution of sources over the map, we do not see disjunct islands of overdensities which attract a majority of sources from the training sample. There are only two neurons (at \([19, 0]\) and \([19, 2]\)) in the bottom right corner of the map to which a significantly higher number of sources fit best.

This discrepancy can be explained by the composition of our data set. From the classical \(\alpha_\mathrm{IR}\)-index we have estimated that roughly \(10\%\) of all sources in the training sample exhibit interesting morphologies -- i.e., feedback cavities, jets and outflows, etc. -- in the observed wavelengths. The remainder of the sources appear as point sources which are hence overrepresented in the sample. Therefore, we cannot expect a perfectly uniform distribution of sources when mapping the images back onto the trained SOM.

\begin{figure}
  \centering
  \resizebox{\hsize}{!}{\includegraphics{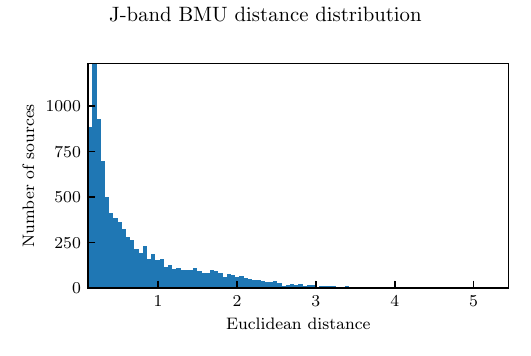}}
  \caption{J-band distribution of BMU distance of each image in the training sample. The successfully trained SOM features a distribution of BMU distances with a peak at very small distances and a tail of large distances representing outliers.}
  \label{fig:bmu_distance_distribution}
\end{figure}

To test the second criterion, we look at the distribution of BMU distances for each image in the sample. With a well trained map, we expect that each image in the training sample is well represented by at least one neuron in the SOM. In reality, however, it is common that the neurons in the SOM are a good representation for the great majority of input images rather than for all. In terms of the BMU distances, this results in a BMU distance distribution that peaks at a very small Euclidean distance. A long tail of large distances shows the presence of outliers in the data that can not be represented by the morphological prototypes learned by the SOM. Figure~\ref{fig:bmu_distance_distribution} shows the distribution of BMU distances for the J-band SOM. We see a distribution peaking at a distance of roughly \(0.1\), steeply declining into a shallow tail of larger distances.

For the NIR maps, this fits well with what we expect from a well-trained SOM, hence, leading us to the conclusion that the training of our NIR maps has been successful. The maps created from images at the longer wavelengths may show different morphological prototypes, but the source distributions of back mapped images onto the SOM show that the images are strongly concentrated along the edges and corners of the maps.

\section{Results}
\label{sec:results}
In the following, we present our results from training three maps, one for each filter for which we have gathered the image cutouts. We created a grid of 400 morphological prototypes for each of the three filters. Furthermore, we have analyzed how the \(\alpha_\mathrm{IR}\)-index and, hence, different observational classes distribute over the SOMs. Therefore, we were able to identify morphological prototypes representative of certain YSO classes in a quantitative and statistically robust way.

\subsection{YSO morphological prototypes}
\label{sec:prototypes}

\begin{figure*}[t]
  \centering
  \includegraphics[width=17cm]{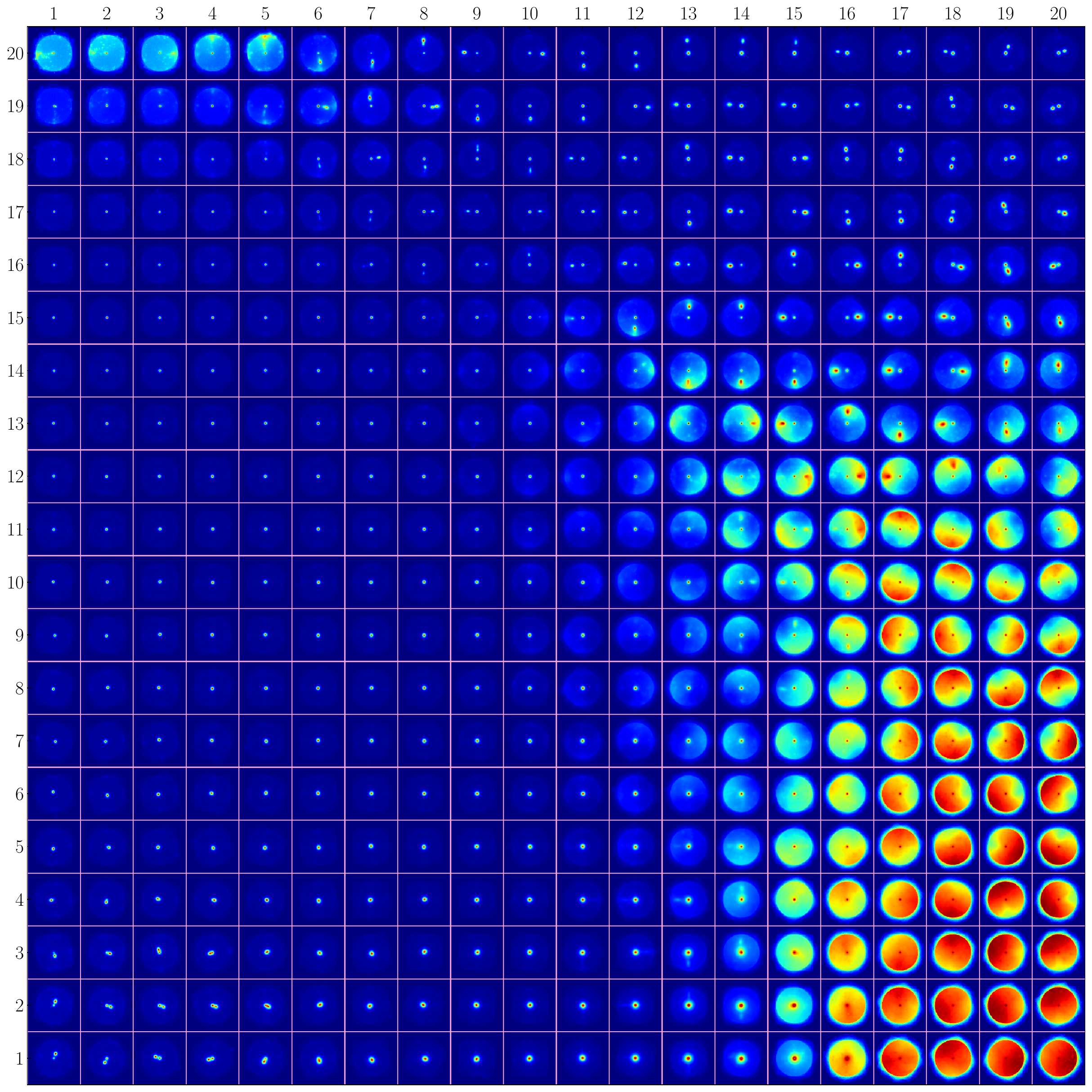}
  \caption{J-band SOM prototypes. The numbers along the left and top edges indicate the coordinates in the latent space created by the Self-Organizing map algorithm, that is, the coordinates of the neurons representing each prototype.}
  \label{fig:J_prototypes_main_text}
\end{figure*}

Figure~\ref{fig:J_prototypes_main_text} shows the morphological prototypes obtained from the J-band image cutouts. The SOMs showing the prototypes in the other NIR bands can be found in Appendix~\ref{sec:appendix_SOMs}. In the following, we will explore the different morphologies learned through the PINK algorithm.

The large region on the left side, ranging from line one at the bottom to line 18 close to the top and reaching from column one to ten, is mostly populated by isolated point sources. The bottom right corner shows prototypes with extended emission from the surrounding interstellar medium. In the J-band, this extended emission can be attributed to the dust cloud, which scatters and reflects the light from the stars in the region. There are also neurons that only partially show extended emission -- e.g. the square of nine neurons centered around line 9, column 19 -- which could indicate that sources that map to these neurons are situated at cloud edges. The map also shows prototypes that appear to have jet-like structures, for instance, in row 20, columns one through six. Finally, we also observed two kinds of visual binaries that we dubbed close and separated. The close binaries are the ones where the two points are connected; that is, their point spread functions are touching, and the source looks elongated with or without a distinctive double peak. These close visual binaries can be found in the bottom left corner of the SOM. Separated binaries show a second isolated point source in the region surrounding the central point source, with neurons showing these prototypes in the top right of the map.

\subsection{YSO observational class distribution}
\label{sec:class_distribution}
To assess how different observational classes are distributed in the SOM embedding space, we calculated for each image in our sample which prototype in the SOM is the most likely to represent the features contained in the input images. These probabilities enabled us to create heatmaps that highlight which observational classes can be found in what area of the SOMs and, hence, give insight into whether there is a link between the \(\alpha_\mathrm{IR}\)-index and the morphology of a YSO. 

\begin{figure*}
  \centering
  \includegraphics[width=17cm]{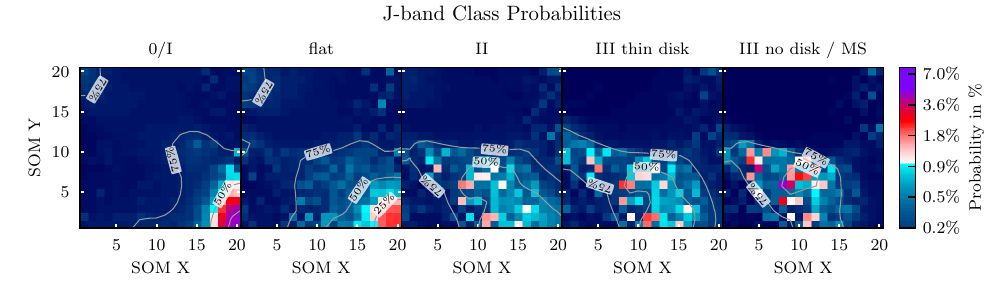}\\
  \includegraphics[width=17cm]{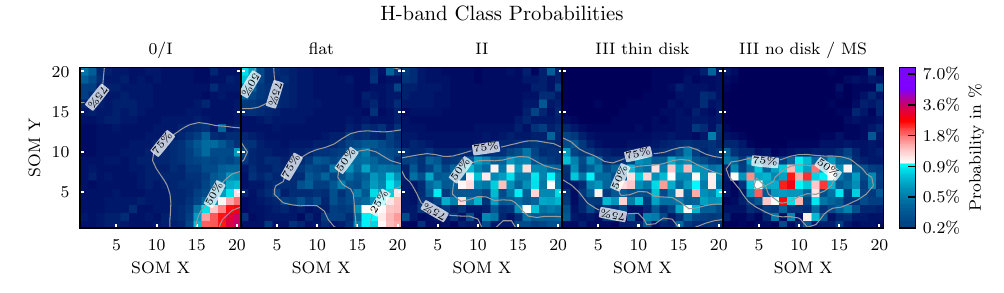}\\
  \includegraphics[width=17cm]{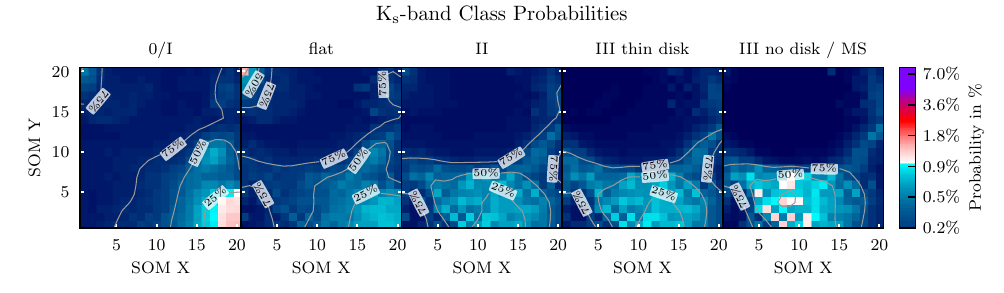}
  \caption{Probability density function for each observational class in the J-, H-, and K\textsubscript{s}-band. These maps highlight the regions on the SOM where we have the highest probability of finding a specific class of YSO. The five panels from left to right show the probabilities per neuron for each class, starting with the youngest on the left and ending with the oldest, most evolved class to the right. The contours in each panel enclose the area within which 25\%, 50\%, and 75\% of all YSOs of a given class are located within the map.}
  \label{fig:class_probability_distribution_NIR}
\end{figure*}

The three rows in Fig.~\ref{fig:class_probability_distribution_NIR} show the probability density functions for each YSO class for each of the three SOMs trained with the images observed in one of the three NIR passbands, where the first, second, and third rows correspond to the J-, H-, and K\textsubscript{S}-band SOMs. Each of the five panels of each row show the most probable area in which the image of a specific class of YSO can be found. That is, each pixel in each heatmap represents the probability that the YSO prototype located at the same x and y coordinates in the SOM represents a YSO of the respective Class. To help guide the eye when interpreting the probability distributions for each Class of YSO we have overplotted contours that enclose the area where we can find 25\%, 50\%, and 75\% of all YSOs of the respective Class in each heatmap. The leftmost panel tells us, that Class 0/I YSOs are best represented by the morphological prototypes in the bottom right corner of the J-band SOM. This is also indicated by the contour lines roughly enclosing the bottom right 4 by 4 block of prototypes, which show that 50\% of all Class 0/I YSOs feature morphologies best represented by these prototypes. This can interpreted as there being a strong correlation between morphologies found in the bottom right prototypes and Class 0/I YSOs. The counter-conclusion is that there is no correlation between Class 0/I YSOS and the morphologies found in the rest of the map. The SOM has learned that this region contains images that exhibit significant emission from the surrounding medium and the central source is barely visible, if at all. This is consistent with what we would expect in the J-band since Class 0/I YSOs are hard to observe at this wavelength due to the massive dust extinction caused by the envelope of these deeply embedded proto-stars.

Similar results were found for the flat-spectrum sources, the second panel from the left of Figure~\ref{fig:class_probability_distribution_NIR}. Even though the highest probability is again in the bottom right corner of the J- and H-band maps, we also see that the distribution extends towards the top left, to the center of the SOM, so that the entire lower right quadrant contains possible morphologies in the SOM. There are additional high probability morphologies in the top left corner, which are most evident in the H- and K\textsubscript{s}-band (see Figures~\ref{fig:class_probability_distribution_NIR} middle and bottom rows), suggesting that the morphologies found in this area are of significance for the flat-spectrum sources. The prototypes in the top left corner show a central source in the middle of each neuron, and a detached, triangle shaped band of emission which appears to be an outflow that starts rather narrowly and expands the further away it gets from the central star.

The more evolved observational classes -- II, III with thin disk, and III no disk / MS star -- are harder to interpret as their probability distributions all appear to be very similar to each other. However, we can tell that they generally avoid the regions that we found for the youngest -- Class 0/I, and flat-spectrum -- sources, where the flat-spectrum sources act as an intermediate stage showing higher levels of probability in the areas for both families of YSOs, less and more evolved.

\begin{figure}
  \centering
  \includegraphics[width=\columnwidth]{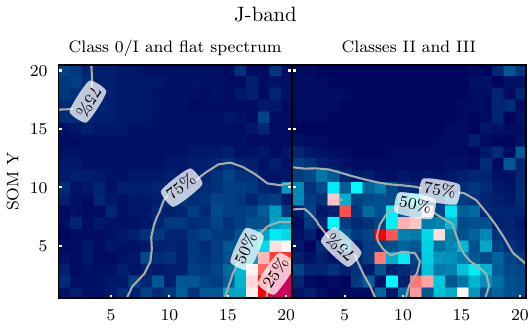}\\
  \includegraphics[width=\columnwidth]{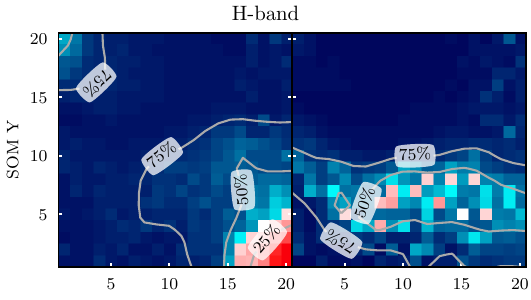}\\
  \includegraphics[width=\columnwidth]{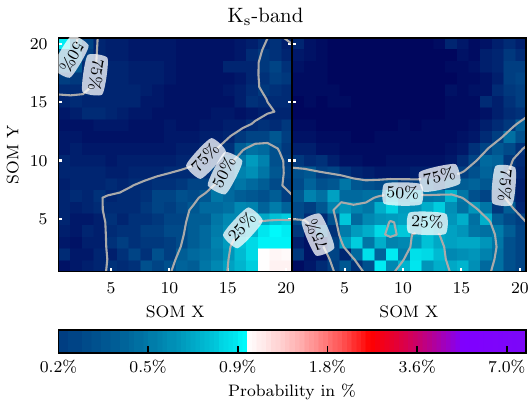}
  \caption{Similar to Fig. \ref{fig:class_probability_distribution_NIR}. Probability density function for YSOs grouped into Class 0/I and flat spectrum sources, and Class II and III YSOs or MS stars.}
  \label{fig:contours_classes_grouped}
\end{figure}

\section{Discussion}
\label{sec:discussion}
We were able to identify certain regions in the SOM that are preferably inhabited by certain YSO classes. However, there are some limitations to what we can achieve with this method, and the data at hand. Thus, the following paragraphs will discuss what we can learn from the trained SOMs, and point out the current limitations. Finally, we will also take a look at how the results from this method could be improved in the future.

With the insights already mentioned above in Sect.~\ref{sec:class_distribution}, we can tell that it is possible to separate the earliest and latest stages of star formation. That is, based on morphology alone we were able to distinguish Class 0/I proto-stars and flat-spectrum sources from older, more evolved, Class II (classical T-Tauri) and Class III YSOs.

\subsection{Proto-stars}
\label{sec:discussion_proto-stars}
First, we take a closer look at the Class 0/I proto-stars. In the SOMs for the NIR filters -- J, H, and K\textsubscript{s} -- we can find them in the bottom right corner. This is the region where YSO prototypes that exhibit strong emission from the surrounding medium are situated. In some cases, the prototypes also show central sources embedded in the medium. This is in line with what we expect for the least evolved, i.e. the youngest YSOs in our sample. At this stage of star formation, we have a central source that is deeply embedded in its proto-stellar envelope.

Although the central object is actively accreting matter from its envelope through an accretion disk, which causes feedback mechanisms to expel a fraction of the in-falling material back into the environment, these outflows have not had enough time to excavate much of the envelope. Thus the jets and outflow cavities are hard, if at all, to detect in the NIR-bands close to the source. The same is true for the central proto-star itself. 

However, outflows driven by Class 0 YSOs have been detected in the NIR- and MIR-bands, most notably the K\textsubscript{s}- and IRAC2-bands, they are often large structures that can reach parsec scales \citep[see e.g., ][]{bally1994, devine1997, mader1999}. The reason why we do not see them in our SOM prototypes is due to the size of the image cutouts that are on the order of \(20\,000\times20\,000\;\mathrm{au}\).

\subsection{Flat-spectrum sources}
\label{sec:discussion_flat-spectrum}
Perhaps the most interesting results from our morphological analysis are the morphological prototypes associated with the flat-spectrum YSOs. The true nature of this class of YSO is still heavily debated in the literature \citep[see e.g., ][]{crapsi2008,dunham2015,tobin2020}. Common theories assume that this type of YSO represents an intermediate evolutionary stage between proto-stars and T-Tauri stars \citep{greene1994,spezzi2011,heiderman2015,furlan2016}, or that the flat-spectrum is not a result of a distinct intermediate stage, but rather an effect of the physical orientation of the YSO with respect to the observer \citep{whitney2003a,whitney2003b,robitaille2006,crapsi2008}.

A flat-spectrum source's most prominent characteristic is the flat slope of their infrared SED (\(\alpha_\mathrm{IR}\approx0\)), which is eponymous to this class of YSO. The most common interpretations of the flat slope are that there is a significant contribution from the central source, but also the accretion disk and envelope. Hence, the superposition of SEDs from the individual components results in the flat shape of the overall SED of the source. Furthermore, this class of YSO shows solid evidence of highly active accretion and feedback processes \citep{spezzi2011}. Moreover, \citet{heiderman2015} also argue in favor of an intermediate stage due to their detection of dense gas in close proximity to the central source of flat spectrum YSOs which also contributes to the flat shape of the SED.

We find that the prototypes associated with this class of YSO are split between sources that appear to be embedded in their surrounding environment as well as sources that show locally confined emission that goes in line with shocked material impacted by jets (see the second panel in each row of Figure~\ref{fig:class_probability_distribution_NIR}). The YSO class probability distributions show that the flat-spectrum sources featuring outflow structures are most prominently present in the H- and K\textsubscript{s}-band observations. This can be explained by strong line emission of molecular Hydrogen (H\textsubscript{2} at \(\lambda=2.12\;\mu\mathrm{m}\)) which is excited when the jet rams into the cloud surrounding these proto-stars \citep{ray2021}. Moreover, jets from YSOs commonly show forbidden Iron lines ([Fe II] at \(\lambda = 1.64\;\mu\mathrm{m}\)) that can be detected in H-band observations \citep{reipurth2001}. This is in good agreement with our results where the areas of the SOMs show a high probability for flat-spectrum sources in the areas populated by prototypes showing outflow signatures. Moreover, the splitting of flat-spectrum sources having both morphologies, those with outflows and jets as well as embedded point sources, corroborates the results of \citet{habel2021}, who find that flat-spectrum sources appear either as point sources or irregular by their categorization.

Since the H\textsubscript{2} and [Fe II] emission trace different components of the outflows \citep[see e.g.,][]{dionatos2018, dionatos2020, melnikov2023}, we would expect this to be reflected in the observations of the YSOs and assume that this is also learned by the SOM. Indeed we find that the prototypes in the top left of the H- and K\textsubscript{s}-band SOMs -- those showing jets structures -- appear to be somewhat narrower in the H-band (sensitive to the [Fe II] line at \(\lambda = 1.64\;\mu\mathrm{m}\)) than they are in the K\textsubscript{s}-band (sensitive to the H\textsubscript{2} line at \(\lambda = 2.12\;\mu\mathrm{m}\)).

As to why the J- and H-band filter SOMs show a lower probability for jets could be explained by extinction effects in the shorter wavelengths. \citet{reipurth2001} mention that there are additional bright lines of [Fe II] observable by the J- and H-band filters, but these lines are often difficult to detect as they are predisposed to extinction in the surrounding ISM. Atomic Hydrogen lines in the NIR (Pa\(\beta\) and Br\(\gamma\)) have also been observed and been brought in connection with YSO jets, yet these lines are only observed in a small fraction of YSO outflows \citep{carattiogaratti2015}.

Moreover, \citet{carattiogaratti2015} point out that extinction effects often introduce an additional constraint to detect atomic jet tracers in the receding lobe of a bipolar outflow. In addition to that, \citet{carattiogaratti2015} also theorize that asymmetric matter distribution in a YSOs envelope, or source multiplicity can prohibit bipolar outflows altogether. This could explain why our prototypes primarily show unipolar outflow structures. 

With this in mind, we would like to note that based on the morphology of the sources alone, we cannot give new insights into the nature of the flat-spectrum YSOs. However, through our quantitative study of several thousand sources, using unsupervised machine learning, we see that YSO outflows have a high probability to be detected in the K\textsubscript{s}-band observations of flat-spectrum sources, even more prominently than for the Class 0/I proto-stars. The lack of jet signatures found by the SOM for the youngest proto-stars can be attributed to the high extinction in the proto-stellar envelope, effectively cloaking these outflow structures from our sight. On the other hand, it is possible that Stage I YSOs with visible outflows at low inclinations, i.e., viewing down into the narrow feedback cavity of the envelope, the source could have been misclassified as a flat spectrum source.

\subsection{Class II and III YSOs}
\label{sec:discussion_class_II_III}
Finally, we are unable to separate the more evolved sources, that is, the Class II T-Tauri stars from the Class III pre-main sequence stars. These objects have already consumed their proto-stellar envelopes, and what remains are at first thick, and as they evolve further, thinner and thinner disks as planets form -- further depleting the dust and gas in the disk. In addition, stellar winds blow away the remaining gas. Thus, the images of these systems show predominantly point sources. This directly results in a degenerate morphology that can not be classified without further information, e.g. flux ratios from different filters, or spectroscopic data. As a consequence, PINK is not able to learn any differences between these classes from the data available to us.

\subsection{Limitations}
\label{sec:discussion_limitations}
Some limitations of using SOMs for morphological classification have already been mentioned in the sections above. Here, we give a summary of said limitations and caveats we have identified.

First to mention is the number and demography of sources in the training sample. From the \(27\,879\) literature YSO candidates \citep{roquette2025} only about a third (\(10\,355\)) are observed by the VISION survey. In general, SOMs are trained on training sets that are at least one to two orders of magnitude larger. To compensate for this, we had to extend the number of training epochs to arrive at maps that were well trained. One possible path to explore in a future paper would be to train PINK on images obtained from theoretical models, see e.g. \citet{whitney1992,whitney1993} or \citet{robitaille2011a}, which would allow a large enough dataset of synthetic YSO images.

A bigger caveat is the composition of morphologies found in the training sample. Out of the \(10\,355\) sources, we were able to calculate \(\alpha_\mathrm{IR}\)-indices for \(\approx 8\,355\) objects. Since interesting morphologies -- e.g. jets, outflow cavities, shocked material -- which are mainly a feature of YSOs in their earliest stages of evolution, are the least common. Hence these morphologies are largely underrepresented in the training sample. Based on the number of sources for which we have derived an observational class (\(\approx 8\,355\)), Class 0/I (723) and flat spectrum sources (628) only account for roughly \(20\%\) of the sources. Taking into account literature estimates for the fraction of the youngest YSOs, e.g. from \citet{grossschedl2019}, this fraction lies at \(12.5\%\) which is even lower than the 20\% given above.

Having a large variety of different morphologies we want the SOM to learn, for instance, outflow cavities with different opening angles, seen from various inclination angles, with or without jets, bipolar or monopolar, becomes an immense challenge when they are rare compared to other, dominant morphologies, e.g., point sources. One way to counter this issue would be to remove all point sources, which could be easily done using the SOM by excluding all sources that map back onto the prototypes representing a point source. However, for our study, this is not an option as this would diminish the size of our training set too much for it to still be a viable sample.

Another limitation we found is the resolution of the observations. To distinguish different morphologies we demand that the defining features of these morphologies can be spatially resolved. With proto-stellar disk sizes ranging from a few dozen to several hundred au, VISTA/VIRCAM -- with a pixel scale of \(1/3\arcsec\;\mathrm{per}\;\mathrm{px}\) (\(0.72\arcsec\;\mathrm{per}\;\mathrm{px}\) after resampling) which at the distance of Orion equates to \(\approx\,150\;\mathrm{au}/\mathrm{px}\) (\(\approx 300\;\mathrm{au}/\mathrm{px}\) after resampling) -- is powerful enough to potentially resolve the proto-stellar disks and outflows, the latter of which are typically significantly larger than the disks. Unfortunately, the resolution of the MIR images from \emph{Spitzer} was not high enough for us to come to a satisfying conclusion.

While training our SOMs we have also identified crowding as a limiting factor impacting the performance of our morphology analysis, especially in the ONC and its immediate neighborhood. To a certain extent, crowding can be managed by choosing an appropriate size of the image cutouts. Ideally, we wish to have only the YSO and its outflows in the image. To include the entire system, especially with large scale outflows, a larger stamp size would be better. However, with increasing image dimensions, more unrelated contaminating sources will be included in the image. Hence, the image dimensions must be carefully weighed between a larger size to encompass much of the outflows and a smaller size to limit contaminating sources that can confuse the SOM algorithm. 

Unfortunately, the image stamp size chosen in our case can not exclude all contaminating sources from the each cutout. We found that slightly less than 60\% contain at least two sources listed in the NEMESIS Catalogue (see Sect.~\ref{sec:data_selection}). In absolute numbers this amounts to 5675, 5874, and 5848 for the J-, H-, and K\textsubscript{S}-band image cutouts, respectively. Out of the multiple source images the ratio between cutouts with only two and those with more than two YSOs is proximately 40\% to 60\%. The SOMs we trained do show prototypes for binary sources, but we did not observe any prototypes containing more than two sources. Even though PINK is a rotation (about the center of the image) and flipping (along all axes running through the image center) invariant implementation of a SOM, it is not invariant to the number of sources or their spatial arrangement inside the image cutout.

To understand how PINK handles cases of multiple sources we consider two general cases. First, an image containing two distinct sources. Since the image cutouts are centered on the YSO coordinates from the input catalog, one source is always in the center, and the second source will be somewhere off center. When PINK determines the best matching unit, it creates copies of the original image at various rotation angles and flipping states. For the SOM update, the algorithm only keeps the copy that best fits the BMU on the SOM. The only parameter PINK can not compensate for in this case is the separation between the two sources, hence there are several similar prototypes showcasing binary sources with differing separation. The second case considers an image with more than two visible sources in the image. Here we have two parameters per off center source that are not marginalized by PINK, the separation and the direction towards where the off center source lies in the image. As a result, these images are similar to noise since the spatial distribution of the sources, with exception of the one located at the image center, is practically random, thus these images are treated by PINK as morphological outliers whose morphology is not learned, i.e., represented in any of the prototypes. This is a double bladed sword. On one side it may be beneficial that PINK somewhat ignores images that suffer from contamination due to crowding. On the other side however, there is no guarantee that PINK actually associates a potential morphological distinctiveness of the source at the image center with a prototype that represents this morphology due to the surrounding contaminating sources in the input image.

\section{Conclusion}
\label{sec:conclusion}

Using a rotation and flipping invariant self-organizing map algorithm, we created a grid of morphological prototypes of young stellar objects for eight near- and mid-infrared bands. This allows us to explore stellar evolution in the earliest stages of a star's life cycle. We found that the best results were produced with the images obtained from the NIR survey VISION.

This study in the Orion A molecular cloud is a preparatory work towards a future spectro-morphological classification scheme for YSOs. The lessons learned in this paper will directly influence future improvements to the method, aiming to use morphological information extracted with PINK -- either from observed of synthetic images -- and combine them with spectral information, e.g., from SEDs and spectra, to develop an improved and more reliable classification for YSOs.

In contrast to a model hypothesis driven analysis, in this study, we have shown that in a data driven quantitative analysis of different YSO morphologies, unsupervised Self-Organizing maps are an effective tool to investigate early stellar evolution, given the sample size is large, and the resolution of the images in the training set is high enough. We found that we can successfully separate less evolved YSOs -- i.e. Class 0/I protostars and flat-spectrum sources -- from the more evolved class II and III YSOs. Moreover, we found that flat-spectrum sources show a high probability for prototypes representing embedded jet-launching sources alike. This is in line with flat-spectrum sources being an intermediate stage between Class 0/I and Class II, as suggested by the literature.

Moreover, along with the typical morphologies we expect from young protostars, we found that PINK has also learned the morphology of close and separated double point sources. Although we did not investigate these morphologies closely, we believe that PINK could be utilized to identify and analyze true and visual binaries in the sample. We have further noticed a large number of resolved galaxies in the VISION image data. Our sample of YSOs in Orion does contain a small number of galaxies, however not enough to impact our YSO morphology prototypes noticeably. Nevertheless, PINK -- if it was trained on VISION NIR images of galaxies -- could prove useful to researchers focusing on galaxies.

\begin{acknowledgements}
  This work is part of the NEMESIS project which has received funding from the European Union's Horizon 2020 research and innovation program under grant number 101004141. G.M. acknowledges support from the János Bolyai Research Scholarship of the Hungarian Academy of Sciences. This research was supported by the International Space Science Institute (ISSI) in Bern, through ISSI International Team project 521 selected in 2021, Revisiting Star Formation in the Era of Big Data (\url{https://teams.issibern.ch/starformation/}). We would also like to thank the anonymous referee for their valuable input from which this paper has greatly benefited.
\end{acknowledgements}

\bibliography{references.bib}

\begin{appendix}

  \section{PINK -- training}%
  \label{sec:appendix_training}
  \paragraph{Similarity metric:}%
  To find the BMU, we have to define a measure of similarity (i.e., a distance measure) to compare the input data to the SOM output. Depending on the data, different distance metrics have been suggested \citep[][]{kohonen2001}; Among them, the Euclidean distance \((L^2-norm)\) is commonly used, but it has been suggested that other metrics may be better suitable in some cases \citep[see e.g., ][]{aly2008,drakopoulos2020}.

  However, we will rely on simple Euclidean distance, as it already yields good results for our first attempt at determining morphological prototypes. Moreover, a detailed analysis of which similarity metric is best suited for YSO morphologies lies beyond the scope of this paper.

  \paragraph{Map size:}%
  Determining the best size for the SOM, i.e., the number of neurons, is essential as the SOM size determines the number of morphological prototypes. A first estimate for the number of neurons is to use the square root of the number of sources in the input data set \(\sqrt{n_\mathrm{sources}}\). However, the set of YSO images we retrieved is dominated by a large number of point sources. A rough estimate of the ratio between images showing point sources and images showing more complex YSOs is roughly 5\%. This is an issue because these 5\% contain the most interesting morphologies, but the map does not have enough space to represent them properly.

  For example, in a training set of approximately \(10\,000\) sources and a SOM size of \(\sqrt{10\,000} = 100\) neurons, each neuron should become a best-matching prototype for 100 sources in the training set. When only 5\% of the training set contains interesting morphologies beyond point sources, there are 5 neurons in the SOM representing all of these morphologies. There are several solutions to this under-representation of interesting morphologies. One is to increase the number of neurons in the SOM.

\begin{figure*}
    \centering
    \includegraphics[width=5.66cm, height=5.66cm]{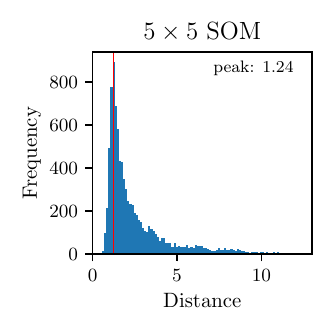}
    \includegraphics[width=5.66cm, height=5.66cm]{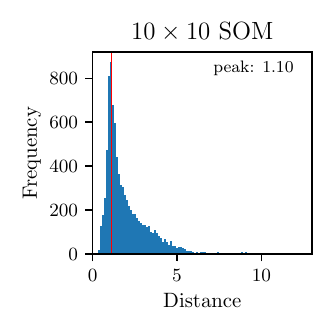}
    \includegraphics[width=5.66cm, height=5.66cm]{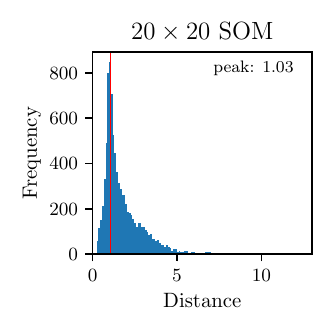}
    \caption{BMU distance histograms for SOM sizes of \(5 \times 5\), \(10 \times 10\), and \(20 \times 20\) neurons. These histograms show the distribution of Euclidean distances between a YSO image and its corresponding best-matching unit for all source images in the training sample. From left to right, the panels show the distribution of BMU distances for increasing SOM sizes. The red vertical line indicates the position of the peak of the distance distribution.}
    \label{fig:distance_histograms}
  \end{figure*}

  We can estimate the optimal size of the SOM by computing the pixel-wise Euclidean distance between each source in the training set and its best matching unit (BMU) in the SOM. If all morphologies are well represented in the SOM, the majority of distances to the BMUs are small. On the other hand, if a large number of diverse morphologies have the same BMUs, the distances to the BMU will increase. Thus, the histogram of BMU distances (see Figure~\ref{fig:distance_histograms}) can be used to measure the optimal SOM size.

  \begin{table}[ht!]
    \caption{Number of image cutouts per filter obtained towards the Orion A molecular cloud.}
    \label{tab:nr_of_images_per_filter}
    \centering
    \begin{tabular}{c c}
      \hline\hline
      Filter & Nr. of cutouts\\
      \hline
      J & \(\phantom{0}9\,855\)\\
      H & \(10\,094\)\\
      K\textsubscript{s} & \(10\,089\)\\
      \hline
    \end{tabular}
  \end{table}

  To determine the optimal SOM size, we aim for a narrow distribution, meaning that the majority of images in the training sample have a well-fitting best-matching unit generated by the SOM. Outliers, which are sources that are not well represented by any of the morphological prototypes in the sOM, will have large distances compared to the majority of sources, located to the far right side of the histogram. Optimally, the tail to the far right of the distance distributions in Figure~\ref{fig:distance_histograms} vanishes when approaching the optimal SOM size. At the same time, the peak of the distribution should move to the left as the bulk of the sources is better represented by the SOM prototypes. In Figure~\ref{fig:distance_histograms}, we compare the three different SOM sizes; the left panel shows the distance distribution for a \(5\times5\) grid, the middle panel for a \(10\times10\), and the tight panel for a \(20\times20\) grid.

  We settled for a SOM size of \(20\times20\) neurons, as the distance distribution peaks at 1.03, which is the lowest of the three map sizes (1.24 for the \(5\times5\), and 1.10 for the \(10\times10\) maps). Moreover, the histogram of the \(20\times20\) grid has a short tail to the right, i.e. few outliers not well represented by the prototypes. Thus, we deemed the \(20\times20\) map is best suited for our experiments.

  Increasing the SOM size further bears the risk of overfitting the model since a larger number of neurons means that there are fewer sources from the training sample map for each prototype. In the extreme case of having just as many neurons, or more, as there are sources in the training set, we would expect to find each input image reflected in exactly one neuron of the map, hence overfitting the model. Yet, there may be some information gained from such a SOM, as it still should group similar sources in similar regions of the map.

  \paragraph{Number of image cutouts:}%
  The total number of image cutouts is slightly different in the individual passbands. This is explained by various differing observation footprints and detector sensitivities. The final number of cutouts per filter is shown below in Table~\ref{tab:nr_of_images_per_filter}.

\section{PINK -- Self Organizing Maps}%
\begin{figure*}[b]
    \centering
    \includegraphics[width=17cm]{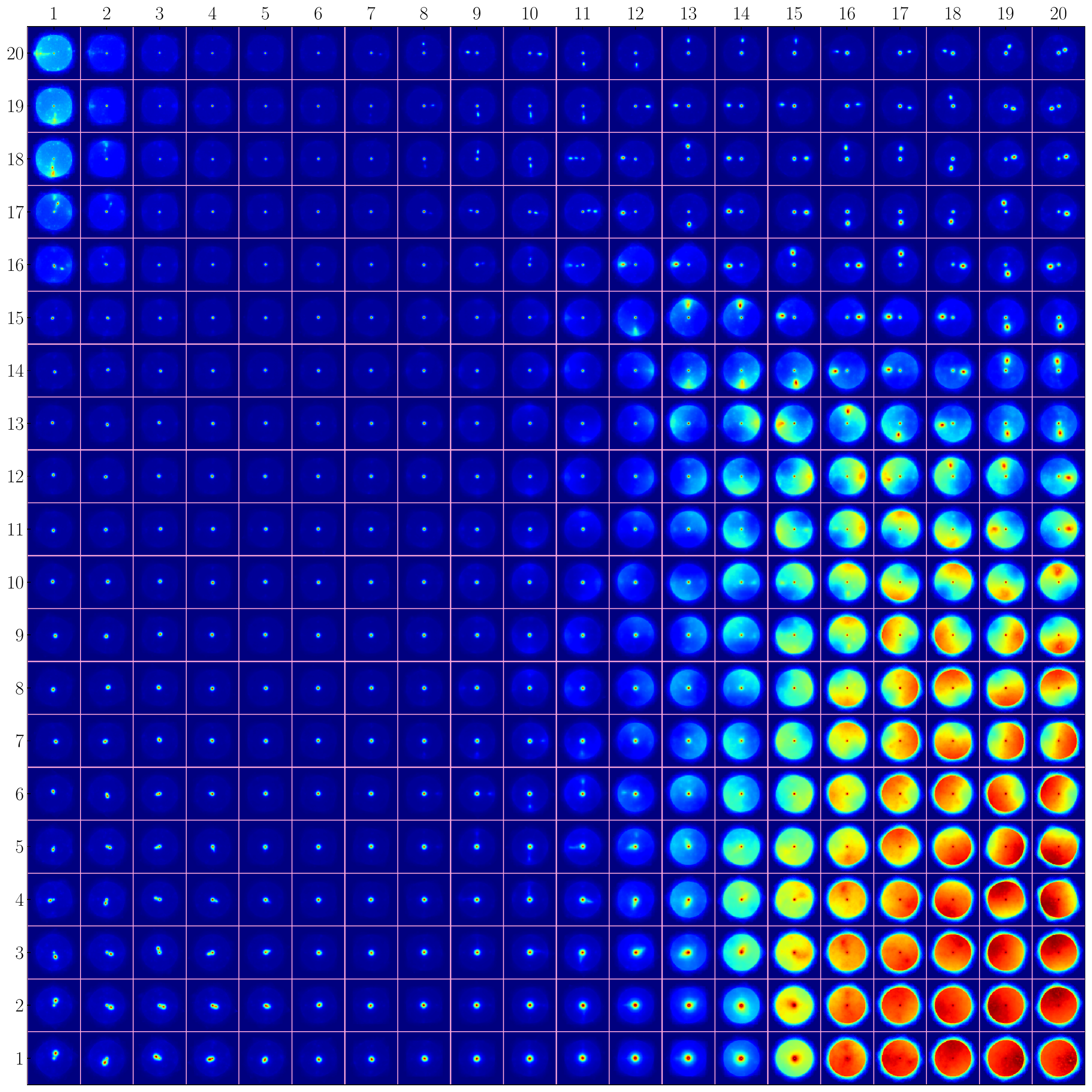}
    \caption{H-band SOM prototypes. Similar to Figure~\ref{fig:J_prototypes_main_text}.}
    \label{fig:H_prototypes}
    \vspace{3.5cm}
  \end{figure*}

\begin{figure*}[htbp]
    \centering
    \includegraphics[width=17cm]{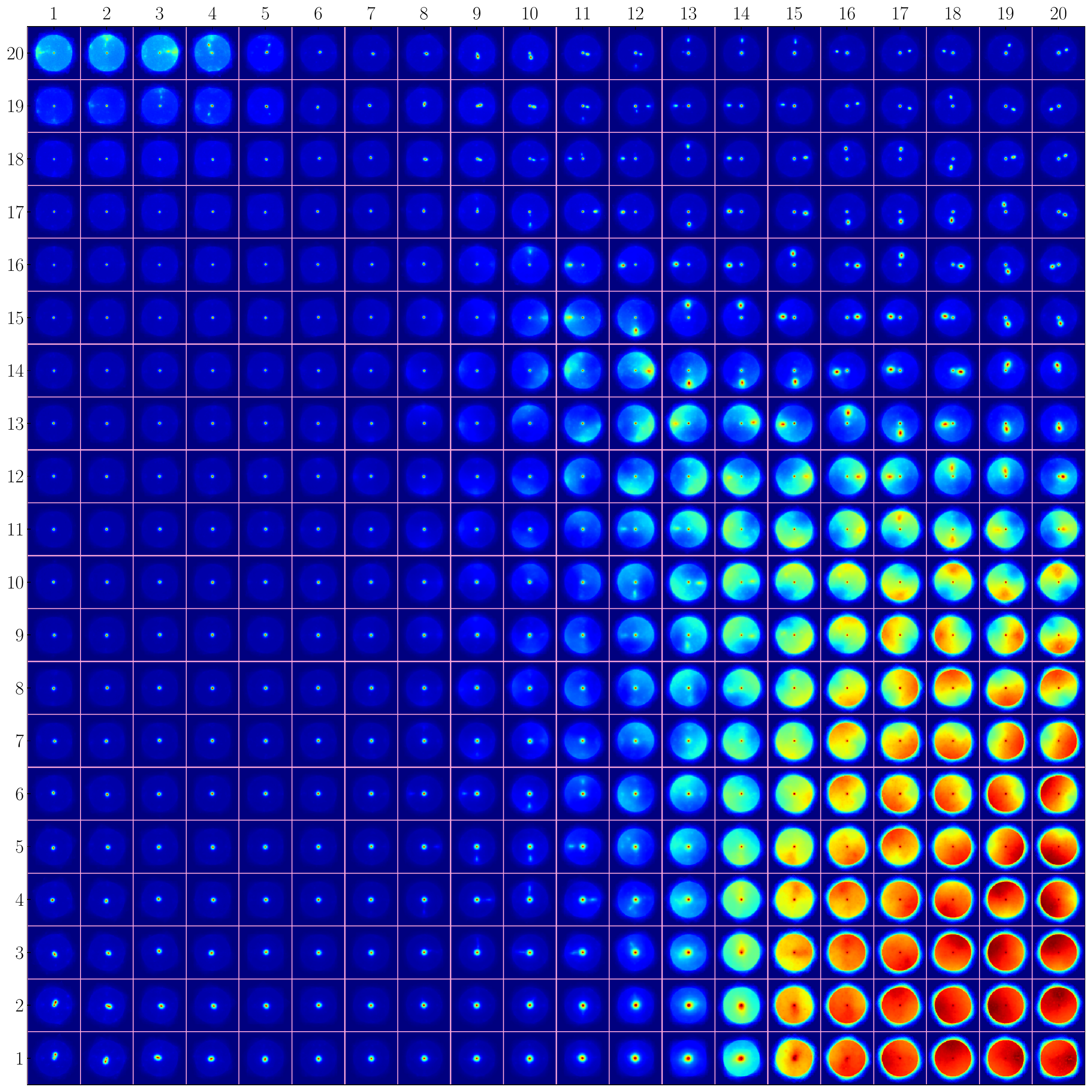}
    \caption{K\textsubscript{s}-band SOM prototypes. Similar to Figure~\ref{fig:J_prototypes_main_text}.}
    \label{fig:Ks_prototypes}
  \end{figure*}

  \label{sec:appendix_SOMs}
  Full-size SOMs for the remaining filters not shown in the main text: H and K\textsubscript{s}
  passbands are shown in Figures~\ref{fig:H_prototypes} and \ref{fig:Ks_prototypes}.
  \FloatBarrier

\end{appendix}

\label{LastPage}
\end{document}